\documentstyle[here,twoside,fleqn,espcrc2,epsf]{article}
\newcommand{\beq}{\begin{equation}}
\newcommand{\enq}{\end{equation}}
\newcommand{\beqa}{\begin{eqnarray}}
\newcommand{\beqast}{\begin{eqnarray*}}
\newcommand{\enqa}{\end{eqnarray}}
\newcommand{\enqast}{\end{eqnarray*}}
\newcommand{\nn}{\nonumber}
\newcommand{\lb}{\label}
\newcommand{\rf}{\ref}
\newcommand{\ct}{\cite}

\newcommand{\lag}{\langle}
\newcommand{\rag}{\rangle}

\newcommand{\al}{\alpha}

\newcommand{\ga}{\gamma}

\newcommand{\ep}{\epsilon}

\newcommand{\la}{\lambda}

\newcommand{\rh}{\rho}
\newcommand{\si}{\sigma}

\newcommand{\ps}{\psi}
\newcommand{\om}{\omega}

\newcommand{\gaga}{$\ga^{(*)}\,\ga^{(*)}\;$} 

\newcommand{\AmS}{{\protect\the\textfont2
a\kern-.1667em\lower.5ex\hbox{M}\kern-.125emS}}
\title{Two-photon reactions at high energies}
\author{
H.G. Dosch\address{Inst. Theor. Physik der Universit\"at, Philosophenweg 16, 
D69120 Heidelberg, Fed. Rep. Germany\\
email: h.g.dosch@thphys.uni-heidelberg.de}\\ }
\begin{document}

\begin{abstract}
We present a short overview over the different contributions to high energy
photon photon scattering and explore the possibilities of tuning the sizes of
the scattered objects by changing the virtuality of the photons. We compare
the experimental data with model calculations. The difficulties with inclusive
charm production are discussed. 
\end{abstract}
\maketitle

\section{Introduction}

The reactions \gaga $\to$ hadrons at high energies can be viewed as scattering
processes of partonic systems with tunable size, since the reaction time of a
highly energetic photon  is short as compared to its dissoziation time  into
partons~\ct{iof69} and the size of the system depends on the virtuality of the
photon. The situation is summarized in figure \rf{gaga1}.

For high virtualities of
both photons not only the parton densities  in the photon can be described well
by perturbative QCD but also the interaction between the small partonic
systems. Therefore we have in \gaga sytems the unique possibility to isolate a
small angle high energy scattering process determined by perturbative QCD. This
investigation is particularly interesting since it can clarify the question of
the strong energy dependence at high photon
virtualities observed at HERA. In deep inelastic scattering the energy
dependence at fixed photon virtuality $Q^2$ is expressed through the
dimensionless Bjorken variable $x=Q^2/(Q^2+W^2)$ where $W$ is the \gaga c.m.
energy. There are currently two principal explanations for the strong
dependence on $1/x$ under discussion. One starts from the usual QCD evolution
using  renormalization group arguments to sum powers of $\alpha_s \,
\log(Q^2/Q_0^2)$. Here the $x$-dependence cannot be predicted but there are
initial conditions at fixed $Q_0^2$ which describe the observed cross sections
very well. The rapid rise of the proton structure function with small $x$ is
then explained by a singularity of the Mellin transform of the DGLAP splitting
function. On the other hand a summation of powers of $\alpha_s \,
\log(W^2/W_0^2)$ first performed by Lipatov and collaborators
\cite{FKL75,BL78} yields   a  power behaviour of the total
cross section  like $(W^2/W_0^2)^{\ep_h}$ where to leading order $\ep_h\approx
\alpha_s 4 \log 2 \approx 0.55 $ even for a small value $\al_s=0.2$.   If this
behaviour is interpreted in the language of Regge trajectories it correponds
to a trajectory with a much higher intercept than the soft pomeron needed to
describe high energy hadron hadron scattering~\ct{DL92}. The summation of 
powers of $\alpha_s \, \log(W^2/W_0^2)$is a  perturbative procedure and the
corresponding pomeron is called the hard pomeron or from the initial of its
authors BFKL-pomeron.  The BFKL formalism has been applied to \gaga scattering
in \ct{BHS97,BRRW99}.

The nice picture seemingly explaining the strong rise
at small $x$ was disturbed by the   calculations in
next-to-leading-order, again by  Lipatov and Fadin \ct{FL98} and an
independent group \ct{CC98}. The contributions to the intercept where as large
as the ones from the leading order and reduced the intercept to a value very
near to one. This clearly indicates that resummations are necessary. Attempts
in this direction \cite{Sal98,CC99} lead an intercept near 1.2. 
\begin{figure}[h]
\leavevmode
\centering
\begin{minipage}{7.5cm}
\epsfxsize7.5cm
\epsffile{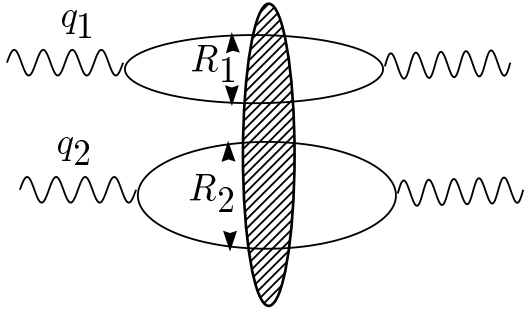}
\centering  
\end{minipage} 
\caption{The diffractive contribution to hadronic \gaga
scattering. The size of thw quark anti-quark system is inversly proportional
to the virtuality: $R_i \sim 1/Q_i = \protect \sqrt{-q_i^2}$.\lb{gaga1}}
\end{figure}

Objections have been brought forward against the
DGLAP evolution, and a
solution~\cite{DL98}  which is also very successful from a  phenomenological
point of view has been proposed, namely the introduction of two pomerons; a
soft one relevant for low $Q^2$ with intercept 1.08 and a hard one important
for high $Q^2$ values with intercept 1.42, i.e. near the leading order  BFKL
intercept.  Since in $\ga^*\,\ga^*$ reactions  with highly virtual photons no
non-perturbative input as the initial conditions of the parton distributions
of the proton are needed this reaction plays a priviliged role for the
clarification of the nature of the energy dependence in perturbative QCD.

The experiments of LEP2 open thus a new window to investigate  diffractive
prozesses, but the kinematic range is not such as to make non-diffractive and
soft diffractive contibutions negligible. We put in this contribution   the
main emphasis on the soft diffractive contributions. They are modelled by an
approach to high energy scattering and non-perturbative QCD which turned out
to be successful in many reactions. 
The total \gaga cross section is in this description given as a superposition 
of the forward scattering amplitude $T(W^2,R_1,z_1,R_2,z_2)$ 
of two colour-singlet quark-antiquark 
pairs~\ct{KD91} with size $R_1$ and $R_2$ respectively  
with the $\rh_\ga$ the quark densities in a photon (see \rf{rho}) as weight
factors
 \beqa 
\lefteqn{\si_{\ga^*ga^*}=\int d^2R_1\,d^2R_2\int_0^1 dz_1\,dz_2
\rh_\ga(Q_1^2, R_1,z_1) }\nn \\ &&
 \rh_\ga(Q_2^2, R_2,z_2)T(W^2,R_1,z_1,R_2,z_2) . 
\lb{1} \enqa 
If in a scattering prozess one object is
small and one large the variation of the amplitude with the size of the small
object can be calculated both in perturbative QCD and in non-perturbative
models. They both predict a variation (apart from logarithms) of the forward
scattering amplitude $\sim R^2$ and hence $\sim 1/Q^2$. Therefore
it is not astonishing that both perturbative and non-perturbative calculations
give reasonable results also outside their domain of strict applicability. If
however both objets are of equal size and small, $R= R_1\approx R_2$, the
behaviour is very different: in perturbative QCD the cross section is
proportinal to $R^2$, in the non-perturbative model however like $R^4$. This
opens the aforementioned separation of the perturbative domain. But even in
this case the perturbative model is useful since it allows to estimate the
kinematical region where non-perturbative effects can be neglected.

As  non-diffractive contributions we have the box diagram ( figure \rf{gaga2}
(a) ) and the reggeon exchange, indicated in figure \rf{gaga2}  (b).

\begin{figure}[ht]
\centering
\begin{minipage}{7.5cm}
\epsfxsize7.5cm
\epsffile{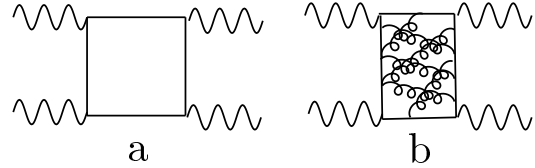}
\centering  
\end{minipage} 
\caption{a: The box diagram giving rise to a fixed pole in the complex $J$
plane. b: A schematic description of the reggeon exchange. \lb{gaga2}}
\end{figure}

\section{Non-diffractive and soft diffractive contributions to \gaga
scattering}

{\em The box diagram} of figure \rf{gaga2} (a) plays a special role. For very
large virtualities of the photons or high quark masses the QCD corrections to
the box diagram are under control and it should not be modified by them in an
essential way. In Regge language the box diagram gives rise to a fixed
double pole at $J=0$ in the angular momentum plane. Its forward amplitude
without any approximations is given in \ct{DDR00}. It decreases with energy
$~\sim \log(W)/W^2$, it is important at moderate energies and large
virtualities. 

{\em The reggeon exchange} has to be added to the fixpole contribution. It is
perhaps the biggest source of uncertainty in the model calculations.  We use
the factorizing expression~\ct{DDR00}: \begin{eqnarray}
&&\hspace{-7mm} \sigma_{\gamma^* \gamma^*}(s,Q_1^2,Q_2^2) =\\ 
&&\hspace{-5mm}4\pi^2\alpha^2{{C}\over{A}}\bigl({{a}\over{Q_1^2+A}}\bigr)^{1-\eta}
\bigl({{A}\over{Q_2^2+A}}\bigr)^{1-\eta}
\bigl({{s}\over{a}})^{-\eta} {\rm nb.}\nonumber
\end{eqnarray}
with $A=0.3$ GeV$^2$, $C=0.38 $,and $\eta=0.45$.
It shows the typical energy behavoiour of the exchange of the reggeon
trajectory.

In the limit of one the photon virtualities $Q^2$ going to zero it 
agrees for $x\geq 0.1$ 
very well with  the hadronic photon structure function related to the DGLAP
evolved  pion
structure function by naive VDM . It might somewhat
overestimate the reggeon contribution at small virtualities: If it is
extrapolated to $\gamma\gamma$ scattering the Regge-contribution to 
$\sigma_{\gamma\gamma}$ is about a factor 3 larger than expected from
factorization and the values obtained in \cite{DL98}. 

{\em The soft diffractive contribution} is evaluated in an  approach
\ct{Nac91} to high energy scattering particularly suited to incorporate
non-perturbative aspects of QCD. The non-perturbative  behaviour of QCD is
treated in the model of the stochastic vacuum \cite{Dos87,DS88}.
It is based on the assumption that the infrared
behaviour of QCD can be approximated by a Gaussian stochastic process in the
gluon field strength. With this assumption one can evaluate the
expectation value of Wegner-Wilson loops and one obtains confinement for
non-Abelian theories. Allthough the model was originally formulated in
Euclidean space-time it can  be extended  to describe the expectation value of
two Wegner-Wilson loops in Minkowski space with lightlike sides
\ct{KD91,DFK94,Dos96}. In this way one obtains the non-perturbative
contribution to the scattering amplitude of two quark-antiquark pairs.

The cross section for scattering of quark-antiquark pairs with sizes $R_1$
and $R_2$ respectively obtained from this model can be approximated to
a reasonable accuracy (10-15 \%)  by the following
simple  factorizing formula:
\beqa
\lefteqn{\si_{dip\,dip}(R_1,R_2) = 0.7 \, \frac{\lag g^2 FF \rag^2}{4 \pi^2}
R_1\,R_2 \times} \nn \\ &&
\hspace{-5mm}\bigg(1-\exp\big(-\frac{R_1}{3.1\,a}\big)\bigg)\,
\bigg(1-\exp\big(-\frac{R_2}{3.1\,a}\big)\bigg)
\lb{siddmsv}
\enqa
where $\lag g^2 FF \rag$ is the gluon condensate \cite{SVZ79}, and $a$ the
correlation length of the gauge invariant gluon field strength correlator, the
numerical values are given below in equation \rf{param}

As can be seen from this equation the model depends essentially on two
typically non-perturbative parameters, which specify the Gaussian process
mentioned above: the strength of the gluon correlator given through the gluon
condensate and $a$, its correlation length. From these on can calculate the
slope of  the linear confining potential \cite{Dos87,DS88}. As it stands the
model leads to  cross sections which are constant with increasing energy. The
parameters of  the model were fine tuned by a fit  to the iso-singlet exchange
part of (anti-)proton-proton scattering at $W = \sqrt{s}$= 20 GeV.  
The resulting parameters which are in the limits of the values determined by
lattice calulations and also leed to the correct value of the slope of the
confining potential are:
\beq
\lag g^2 FF \rag = 2.49\; {\rm GeV}^4 \quad a=0.35\;{\rm fm}
\lb{param} \enq

The
phenomenologically observed  increase with energy of hadronic total cross
sections like $s^{(\alpha_P -1)}$ with $\alpha_P \approx 1.08$ \cite{DL92} can
be incorporated in  two ways: either one lets the radius of the hadrons
increase with $s$  \cite{DFK94,FP97,BN99,PF00}, or one takes the model as a
determination of the  Regge residue and adds the Regge-like increase with
energy by a factor  $(W^2/W_0^2)^{(\alpha_P-1)}$ with $W_0 = 20{\rm
GeV}$. These two approaches  give very similar results, and we adopt the
latter in this paper as it is the  more convenient in the present context.

Whereas hadron-hadron scattering and soft electroproduction processes (i.e. 
those with low photon virtuality $Q^2$ ) can be described  very well in this 
way, this is not the case  for hard electroproduction
processes where the energy ($1/x$) dependence  is much stronger than indicated
by the soft non-perturbative pomeron with intercept $\al_P =1.08$. As discussed
in the Introduction the occurence of a second (hard) pomeron as proposed in
\cite{DL98} can explain the data in a consistent way. This two  pomeron
approach was adapted to the MSV model in \cite{Rue99} and very  successfully
tested for the electro- and photoproduction of vector mesons and, more
relevantly here, for the proton structure function over a wide range of $x$
and $Q^2$. As in \cite{DL98} it was found that the soft-pomeron contribution
to $F_2$, after an initial increase with increasing $Q^2$, has a broad maximum
in the region of 5 GeV$^2$ and then decreases as $Q^2$ increases further i.e.
it exhibits higher-twist like behaviour. In the context of the present model 
this is a consequence of the decreasing interaction strength with decreasing 
dipole size.

It is worth recalling the salient features of this version of the two-pomeron
model which have been slightly modified for the theoretical calculations
presented here, to illustrate the distinction between the soft and the hard
pomeron  in dipole-dipole scattering. In \cite{Rue99} it was assumed that all
dipole amplitudes in which both dipoles are larger than the correlation length
$a=0.35$ fm are dominated by the soft pomeron, and the energy dependence
therefore given by $(W^2/W_0^2)^{\ep_{soft}}$ with $W_0 = 20$ GeV and
$\ep_{soft} =0.08 $. This ensures that the hard pomeron has essentially no
impact on purely hadronic scattering. If at least one of the dipoles is
smaller than $a=0.35$ fm then the energy dependence  is replaced by
$(W^2/W_0^2)^{\ep_{hard}}$ with $\ep_{hard} =0.42 $ . It turned out that the
model overestimated the non-perturbative contribution of very small  dipoles
so it was put to zero if either of the dipoles is less than 0.16 fm. With only
four parameters it was possible to obtain a good description of data for the
proton structure function and for the electroproduction of vector mesons
without noticeably affecting earlier fits to hadron-hadron scattering. 

We apply this two-pomeron model to the evaluation of the
$\gamma^{(*)}\gamma^{(*)}$ cross sections. It should be noted that the 
simple factorisation formula $\sigma_{\gamma\gamma} = \sigma_{\gamma p}^2/
\sigma_{p p}$ is no longer applicable in the two-pomeron situation.

The considerations outlined briefly above lead to a model for the
scattering of quark-diquark dipoles on each other approximated by \rf{siddmsv}.
In order to relate it to $\gamma^{(*)}\gamma^{(*)}$ interactions we have
according to \rf{1} to introduce the photon wave function. 

The quark-antiquark density  in a photon with virtuality $Q^2=-q_\ga^2$
and  helicity $\la$ is to lowest order perturbation theory given by:
\beqa 
\lefteqn{\rho_\ga(\la=0,Q^2,R,z) =}\nn \\ && \hspace{-5mm}
\frac{2 N_c \al}{\pi^2} \hat e_f^2Q^2 z^2(1-z)^2,K_0(\ep R)^2
\nn \\ 
\lefteqn{\rho_\ga(\la=\pm 1,Q^2,R,z) =
\frac{2 N_c \al}{2 \pi^2} \hat e_f^2 \times} \nn \\ 
&&\hspace{-5mm}
\bigg(\big(z^2+(1-z)^2\big) \ep^2 K_1(\ep R)^2
+ m_f^2K_0(\ep R)^2\bigg).
\label{rho}
\enqa
Here $\hat e_f$ is the charge of the quark in units of the elementary charge.
i.e $\pm\textstyle{\frac{1}{3}},\; \pm\textstyle{\frac{2}{3}}$, $m_f$ is
the Lagrangian quark mass; $\la=0$ indicates a longitudinal, $\la=\pm 1$ a
transverse photon. The singularity of the Bessel functions at $R=0$ does not
cause any problems, since for the evaluation of observable amplitudes the
density is multiplied by the dipole cross section and the latter is proportional to 
$R^2$ at small values of $R$ , see \rf{1} and \rf{siddmsv}.
\begin{figure}[ht]
\centering
\begin{minipage}{7.5cm}
\epsfxsize7.5cm
\epsffile{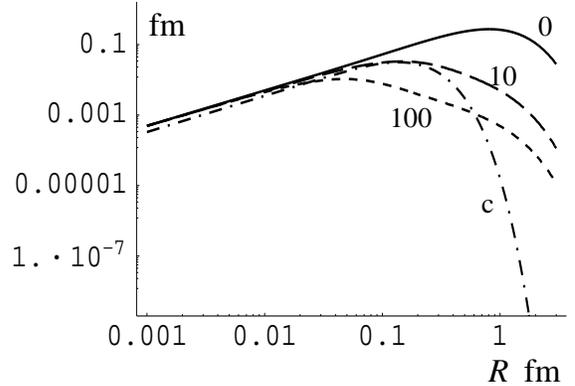}
\centering  
\end{minipage} 
\caption{The quantity $1000/\al  R^3 \int_0^1 dz \rh_\ga(\la=1,Q^2< R,z)$ as
function of $R$ for different values of $Q^2$ for light quarks, solid
$\Leftrightarrow \, Q^2=0$, long dashes $\Leftrightarrow \,Q^2=10$ GeV$2$,
short dashes $\Leftrightarrow \,Q^2=100$ GeV$2$, and for charmed quarks,
$Q^2=0\,\Leftrightarrow $  dot-dashed line.  \lb{gaga3}}  \end{figure}
From \ref{rho} we see that the scale is set by
$\sqrt{z(1-z)Q^2+m_f^2}$.   For longitudinal photons the factor $z^2(1-z)^2$ in
the density $\rh_\ga$  implies that the main contribution comes from the region
$z\approx\textstyle{\frac{1}{2}}$ and the relevant scale is $Q^2/4 +m_f^2$.
For transverse photons however  the endpoints $z=0,\; z=1$ are not suppressed
and a transverse photon might leak into the non-perturbative region even if the
virtuality is quite high. This can be seen directly from figure \ref{gaga3}.

The expressions \rf{rho} are reliable for large values of $Q^2$ and/or large
quark masses (i.e. $c$ and $b$). For small values of $Q^2$ and light quarks the
distance of the quark antiquark pair increases and confinement effects will
become important. 
Often 
vector meson dominance is applied, another possibility is to introduce a
constituent quark mass as infrared regularisator in the photon wave function. 
This approach will be justified in the following, it allows a very economical
description of the photon wave function at low virtualities, which
interpolates smoothly to high virtualities and even takes care of the ``hard
part of the photon'' at low virtualities. This is especially relevant for
photon-photon interactions.

Let us begin with a model  investigation \ct{DGP98} with scalar photons and
quarks. In such a case the ``photon''-wave function at high virtualities is
given by: 
\beqa
 \tilde \ps_\ga(\vec k_\perp) &=& \frac{1}{\vec
k_\perp\,^2 +z(1-z) Q^2 + m_f^2}\nn \\ 
\psi_\ga(\vec R) &=& \frac{1}{2
\pi}K_0(\sqrt{z(1-z)Q^2+m_f^2}\, R)
\lb{mod} \enqa
 For {low $Q^2$}  we expect  confinement to modify these
perturbative expressions considerably.

The structure of \rf{mod} is the same as that of a 
nonrelativistic 
Greens function for the relative motion of a free two body system with reduced
mass $m$  \beq
G_0(\vec R, 0, M) =\frac{m}{\pi} K_0(\sqrt{2 m
M}\,|\vec R \,|),
\lb{free}
\enq
where $mM = - mE $ stands for the virtuality $Q^2$.

In order to impose confinement we go from the free particle state to a system
hold togeter by a harmonic oscillator potential. As has been pointed out in
reference \ct{SVZ79} the harmonic oscillator is a very useful  model for
QCD: it shows both confinement and asymptotic freedom. We therefore investigate
the effects of confinement by comparing the free Greens function \rf{free}
with that of the full harmonic oscillator  
\beq
G_H(\vec R, 0, M) =\sum_{n_1,n_2} \frac{\psi_{\vec n}(\vec
R) \psi_{\vec n}(\vec R)}{(n_1+n_2+1)\om+ M}
\lb{ho}
\enq
which can be calculated easily.

As has been shown in \ct{DGP98}  a free Greens function with a $M$-dependent 
shift: $M \to M + s(M)$ yields an excellent fit to the exact Greens function
\rf{ho}  for  all $M\geq 0$.
We transfer this procedure to QCD by performing 
 a $Q^2$-dependent shift of the flavour mass $m_f \to m_f+ m(Q^2)$ in the
perturbative photon wave function.
The numerical value of the shift $m(Q^2)$ can be fixed by a fit to the
phenomenologically known vector-current two-point  function and thus 
no new parameter is introduced. The following linear parametrizations can be used
\begin{eqnarray} \label{mass}
m_{u,d} \hspace{-3mm}&= &\hspace{-3mm} \left
\{ \begin{array}{r@{\quad:\quad}l} m_0 \,(1-\frac{Q^2}{1.05}) &
Q^2 \le 1.05\\  0& Q^2 \ge 1.05\; \end{array} \right. \\ 
m_{s} & = & \hspace{-3mm}\left\{
\begin{array}{r@{\quad:\quad}l} 0.15 + 0.16\,(1-\frac{Q^2}{1.6})
& Q^2 \le 1.6\;\\  0.15 & Q^2 \ge 1.6\; \end{array} \right. \nonumber\\ m_c &
= & \!\!1.3 \nonumber \end{eqnarray}

The parameter $m_0$ for the $u,d$ quarks was found to be $m_0 = 0.21 \pm
0.015$ GeV.We see that above 1.05 for light quarks and 1.6 for strange quarks
the current mass can be used. For charmed quarks the perturbative expression as
it stands can be taken. 

The densities for transverse photons integrated over the longitudinal momentum fraction $z$
and multiplied by 1000 $R^3$ are shown in figure \rf{gaga3}. The longitudinal
photons are more concentrated at small separations $R$ as explained above.

\section{Results}
\subsection{Total cross sections~\ct{DDR00}}
\begin{figure}[ht]
\centering
\begin{minipage}{7.5cm}
\epsfxsize7.5cm
\epsffile{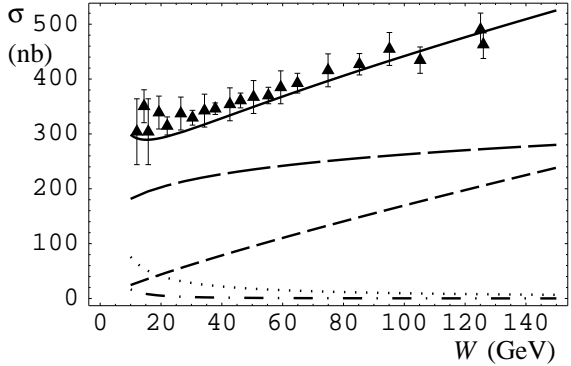}
\centering  
\end{minipage} 
\caption{Cross section (nb) for $\gamma\gamma$ scattering as function of the
c.m. energy $W$ (GeV) compared with L3 data.
L3: Triangles \protect \cite{L3}
The solid curve is our model without adjusted parameters. It consists of the
following contributions: soft pomeron $\Leftrightarrow$ long dashes; hard
pomeron $\Leftrightarrow$ short dashes; fixed pole (box) $\Leftrightarrow$
dot-dashes; reggeon $\Leftrightarrow$ dots.\lb{sigaga0}} \end{figure}
\begin{figure}[h]
\centering
\begin{minipage}{7.5cm}
\epsfxsize7.5cm
\epsffile{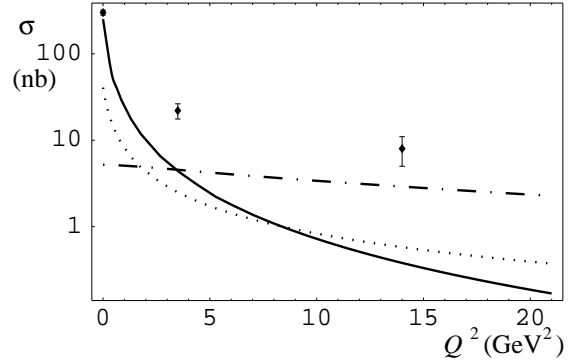}
\centering  
\end{minipage} 
\caption{
Cross section (nb) for $\gamma^*\gamma^*$ scattering as function of
the average photon virtuality $\lag Q^2 \rag = \lag Q_1^2 \rag=\lag Q_2^2
\rag$ (GeV$^2$) a c.m. energy $W \approx 20$ GeV compared with L3 data. Symbols
as in figure \protect \rf{sigaga0} \lb{sigagaqsq} }
\end{figure}

We first discuss the results for the total cross section of real $\ga \,\ga$
scattering. Using the mass $m_q=0.21$ GeV $ m_s=0.31 $ GeV $m_c=1.3$ GeV we
obtain the cross sections given in figure \rf{sigaga0}. The experimental
cross section  shows an increase with energy which is distinctly stronger than
for hadron hadron scattering. Our model reproduces this result and it can be
explained very naturally. The singularity of the photon density $\rh_\ga$ at
$R=0$ which reflects the pointlike coupling of the thoton to the
quark-antiquark pair gives in $\ga \,\ga$ scattering a stronger weight to
smaller dipoles in hadron hadron scattering. Since 
small dipoles couple to the hard pomeron which has  a considerably larger
intercept we obtain an overall energy dependence which is stronger than for
hadrons. The contribution of the box diagram (dot-dashes in figure
\rf{sigaga0})  and the reggeon exchange (dots) are of minor importance above
$W=30$ GeV.  As expected from the discussion in the introduction the purely
non-perturbative model decreases much to fast with virtuality if both photons
have approximatively equal values for $Q^2$.  This can be seen in figure
\rf{sigagaqsq} where the $Q^2$ dependence of the total \gaga cross section is
diplayed as a function of $Q^2=\lag Q_1^2\rag=\lag Q_2^2\rag$ at the fixed c.m.
energy $W \approx 20 $ GeV; at $Q^2=14$ GeV$^2$ the non-perturbative
contribution is negligible. The energy dependence of the \gaga cross section
at $Q^2=3.5$ and 14 GeV$^2$ is shown in figure \rf{sigaga3514}. If the
theoretical non-perturbative and non-diffractive contributions are subtracted
from the experimental values  the resulting energy
dependence can be very well described by a power behaviour $(W^2)^\ep$ with
$\ep = 0.3 \dots 0.4$. 
\begin{figure}[h]
\centering
\begin{minipage}{7.5cm}
\epsfxsize7.5cm
\epsffile{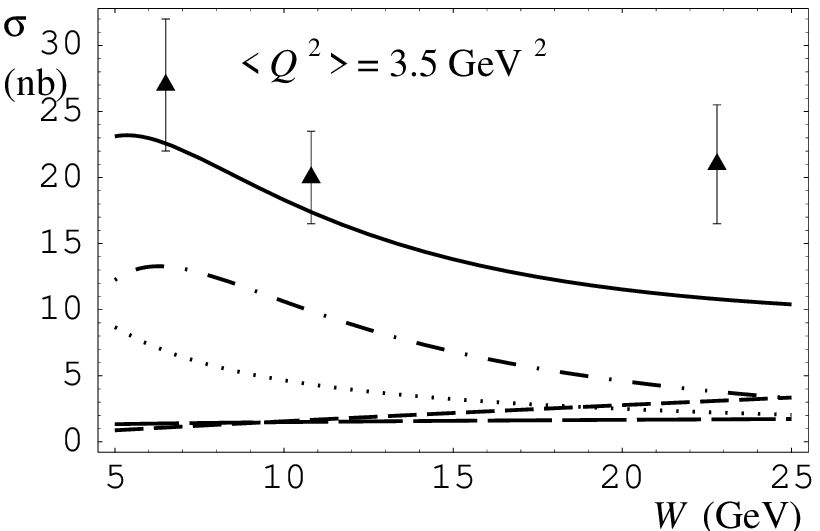}
\centering  
\end{minipage}
\begin{minipage}{7.5cm}
\epsfxsize7.5cm
\epsffile{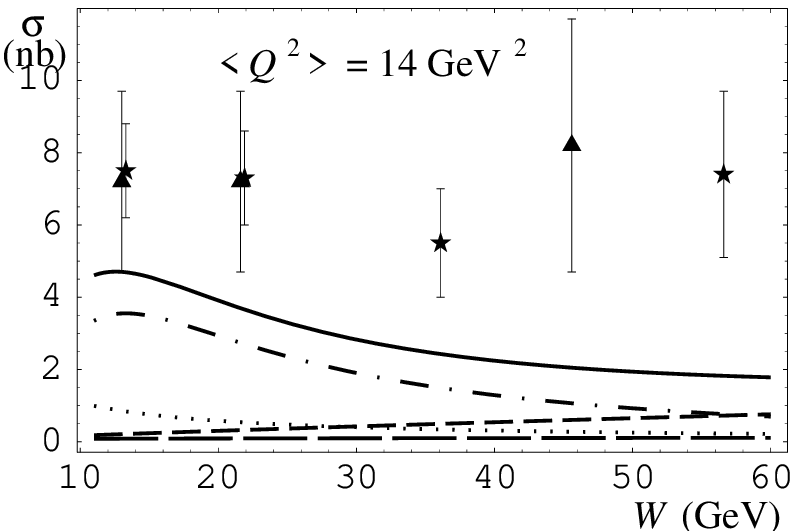}
\centering  
\end{minipage} 
\caption{ Cross section (nb) for $\gamma\gamma$ scattering as function of the
c.m. energy $W$ (GeV) compared with L3 data for $\lag Q^2 \rag = 3.5$ and 14
GeV$^2$ respectively. Symbols as in figure \protect
\rf{sigaga0} \lb{sigaga3514} }\end{figure}

\subsection{Photon structure functions~\ct{DDR00}}

In a photon strucure function one photon is real and the other one has a
virtuality $Q^2 \neq 0$. The structure function $F_2^\ga$ is defined as 
\beq
F_2^\ga(x,Q^2) = \frac{Q^2}{4 \pi^2 \al^2} \si^{\ga^*\ga}(W,Q^2)
\enq
with $x=Q^2/(W^2+Q^2)$.
As
mentioned in the introduction in that case both perturbative and
non-perturbative QCD give the same $Q^2$ behaviour and we see indeed in
figure \rf{modstru} that the non-perturbative model describes the data very
well up to $Q^2=20$ GeV$^2$. In order to extrapolate nontrivially to $Q^2=0$ a
modified structure function  $\tilde F_2^\gamma= \frac{Q^2+0.6 {\rm
GeV}^2}{Q^2}\, F_2^\gamma(x=\frac{Q^2}{W^2+Q^2},Q^2)$ has been displayed and
compared with the correpondingly multiplied experimental data from different
experiments. \begin{figure}[ht] \centering
\begin{minipage}{7.5cm} \epsfxsize7.5cm \epsffile{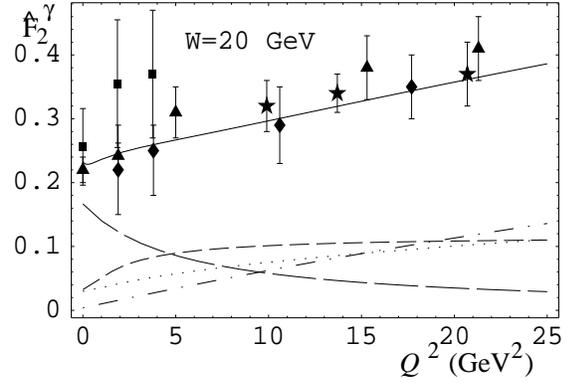}
\centering  
\end{minipage} 
\caption{The modified photon structure function $\tilde
F_2^\gamma=
(Q^2+0.6 {\rm GeV}^2)/Q^2\,F_2^\gamma(x=Q^2/(W^2+Q^2),Q^2)$ as function of
$Q^2$ for $W \approx 20$ GeV. 
The data~\cite{L3,OPAL,ALE} are: Triangles $\Leftrightarrow$ L3 ; 
Diamonds,Boxes  $\Leftrightarrow$ OPAL ; Stars $\Leftrightarrow$ ALEPH 
The solid curve is our fit without
adjusted parameters. For the different contributions see caption of figure
\protect \rf{sigaga0}.
\lb{modstru}} 
\end{figure}
\begin{figure}[hb] \centering
\begin{minipage}{7.5cm}
\epsfxsize7.5cm
\epsffile{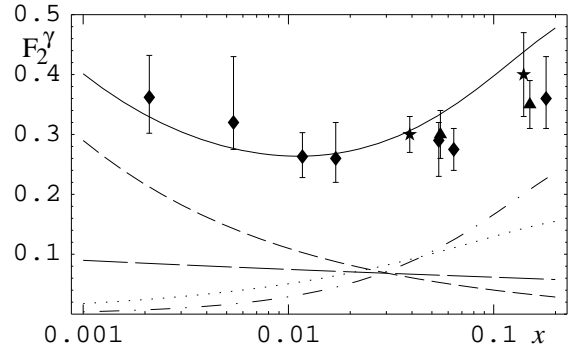}
\centering  
\end{minipage}
\caption{The photon structure function $F_2^\gamma(x,Q^2)$ as function of
$x=\frac{Q^2}{W^2+Q^2}$ for $Q^2 = 10$ GeV$^2$ . 
The data~\cite{L3,OPAL,ALE} are: Triangles $\Leftrightarrow$ L3 ; 
Diamonds  $\Leftrightarrow$ OPAL ; Stars $\Leftrightarrow$ ALEPH 
The solid curve is our fit without
adjusted parameters. For the different contributions see caption of figure
\protect \rf{sigaga0}. \lb{gastru10}}
\end{figure}
In figure \rf{gastru10}  the photon structure function for $Q^2=10$ GeV$^2$ is
shown. The agreement between theory and experiment is similarly good for other
virtualities. 
\clearpage

\subsection{The reaction $\ga^{(*)} \, \ga \to c \bar c \, X$}
\begin{figure}[h]
\centering
\begin{minipage}{7.5cm}
\epsfxsize7.5cm
\epsffile{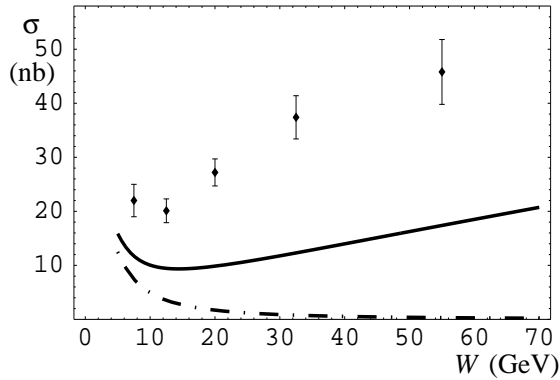}
\centering  
\end{minipage}
\caption{The cross section for the reaction $\ga \, \ga \to c \bar c \, X$ .
The solid line is the direct contribution to the structure function as
predicted by the model from a picture correponding to figure \protect
\rf{gaga1} and the box digram with charmed internal quark lines (dot-dashed
line);
data points are from \protect
\ct{L3C00}. \lb{sigach} }\end{figure} 
New measurements from L3~\ct{L3C00} and OPAL~\ct{OPALC99}  yield
data for inclusive charm production. In figure \rf{sigach} the total cross
section for the reaction $\ga \,\ga \to c\bar c \,X$ is given. We have
calculated diffractive charm production according to the physical picture of
figure \rf{gaga1} where the quarks in the loop of photon 1 are charmed, in loop
2 all flavours are summed over and  correspondingly with 1 and 2
interchanged; for the box diagram only charmed quarks contribute. We see that
the theoretical values for the direct diffractive and non-diffractive
production underestimate the data~\ct{L3C00} by more than a factor two. 
\begin{figure}[h]
\begin{center}
\begin{minipage}{7.5cm}
\epsfxsize7.5cm
\epsffile{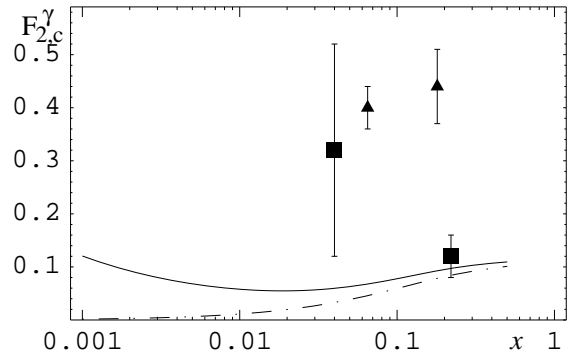}
\end{minipage}
\end{center}
\caption{The photon structure function $F^\ga_{2,c}$ at $\lag Q^2\rag=20$
GeV$^2$. The solid line is the direct contribution to the structure function as
predicted by the model from a picture correponding to figure \protect
\rf{gaga1} and the box digram with charmed internal quark lines (dot-dashed
line). The boxes are the data for the charm structure function 
\protect \ct{OPALC99}, the triangles are  the results for the full
structure function at the same average $Q^2$. \lb{gastruch} } \end{figure}  
The same
happens for the case if one photon is virtual. In figure  \rf{gastruch} the
charm contribution to the photon structure function is displayed for $\lag Q^2
\rag = 20$ GeV$^2$. The central value of the experimental point~\ct{OPALC99} at
$\lag x \rag =0.02 $ is again far above the theoretical value.

\begin{figure}[h]
\centering
\begin{minipage}{7.5cm}
\epsfxsize7.5cm
\epsffile{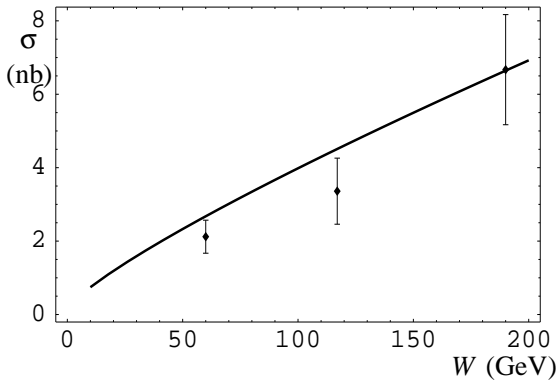}
\centering  
\end{minipage}
\caption{Cross section for the reaction $\ga^* \,p \to c \bar c \, X$. The
solid curve is our model for the direct contribution  for photons with  $Q^2=10$
GeV$^2$. The data points are from \protect \ct{Zeus} \lb{sigapch}} \end{figure}

In principle this could indicate a failure of the model prediction for the
cross section \rf{siddmsv}  for the case that one object is small and one
large, but this is unlikely for the following reasons.\\
 1) The charm structure
function of the proton~\ct{Zeus} agrees very well with the predictions of the
model as can be seen from figure~\rf{sigapch} and here we have the same
situation of one object large and one small.\\
2) The full photon
structure function at $Q^2=10$ GeV$^2$ is correctly predicted by the
model, as can be seen from figure \rf{gastru10}. At $Q^2=10$ GeV$^2$ 
the light quark density in a photon  peaks at about the
same quark-antiquark distance as a charmed pair in a  real photon (see
figure \rf{gaga3}).  

So we would like to conclude that there must be other
mechanisms of charm production but refrain from speculations until the
experimental situation especially of the charmed photon structure function is
clearer. 
\section*{Acknowledgements}
I thank A. Donnachie, A. Hebecker and M. Rueter for many discussions. I also
want to thank S. Narison for having provided again  the familiar and
stimulating atmosphere at the QCD conference 2000 in Montpellier. 

\begin{thebibliography}{10}

\bibitem{Iof69}
B.~L. Ioffe
\newblock{\em Phys. Lett.} B30:123 (1969) 

\bibitem{FKL75}
V.~S. Fadin, E.~A. Kuraev, and L.~N. Lipatov.
\newblock {\em Phys. Lett.}, B60:50, 1975.

\bibitem{BL78}
Ya~Ya Balitskii and L~N Lipatov.
\newblock {\em Soviet Journal of Nuclear Physics}, 28:822, 1978.

\bibitem{DL92}
A~Donnachie and P~V Landshoff.
\newblock {\em Physics Letters}, B296:227, 1992.

\bibitem{BHS97}
J~Brodsky, F~Hautmann, and D~E Soper.
\newblock {\em Physical Review}, D56:6957, 1997.

\bibitem{BRRW99}
M. Boonekamp, A.~De Roeck, Ch. Royon, and S. Wallon.
\newblock {\em Nucl. Phys.}, B555:540, 1999.

\bibitem{FL98}
S~Fadin and L~N Lipatov.
\newblock {\em Physics Letters}, B429:127, 1998.

\bibitem{CC98}
G~Camicin and M.~Ciafaloni.
\newblock {\em Physics Letters}, B430:349, 1998.

\bibitem{Sal98}
G.~P. Salam.
\newblock {\em JHEP}, 07:019, 1998.

\bibitem{CC99}
M.~Ciafaloni and D.~Colferai.
\newblock {\em Nucl. Phys.}, B538:187, 1999.

\bibitem{DL98}
A~Donnachie and P~V Landshoff.
\newblock {\em Physics Letters}, B437:408, 1998.

\bibitem{KD91}
A~{Kr\"amer} and H~G Dosch.

\newblock {\em Phys. Lett.}, B272:114, 1991.

\bibitem{DDR00}
A.~Donnachie, H.~G. Dosch, and M.~Rueter.
\newblock {\em Eur. Phys. J.}, C13:141, 2000.

\bibitem{Nac91}
O~Nachtmann.
\newblock {\em Annals Phys.}, 209:436, 1991.

\bibitem{Dos87}
H~G Dosch.
\newblock {\em Phys. Lett.}, 190B:177, 1987.

\bibitem{DS88}
H~G Dosch and Yu~A Simonov.
\newblock {\em Phys. Lett.}, 205B:339, 1988.

\bibitem{DFK94}
H~G Dosch, Erasmo Ferreira, and A~Kr{\"a}mer.
\newblock {\em Phys. Rev.}, D50:1992, 1994.

\bibitem{Dos96}
H~G Dosch.
\newblock In E.~Ferreira et~al., editor, {\em Hadron Physics 96}. World
  Scientific, 1996.

\bibitem{SVZ79}
M~A Shifman, A~I Vainshtein, and V~I Zakharov.
\newblock {\em Nucl. Phys.}, B147:385, 1979.

\bibitem{FP97}
Erasmo Ferreira and Flavio Pereira.
\newblock {\em Phys. Rev.}, D56:179--183, 1997.

\bibitem{BN99}
E~R Berger and O~Nachtmann.
\newblock {\em Eur. Phys. J.}, C7:459, 1999.

\bibitem{PF00}
Flavio Pereira and Erasmo Ferreira.
\newblock {\em Phys. Rev.}, D61:077507, 2000.

\bibitem{Rue99}
Michael Rueter.
\newblock {\em Eur. Phys. J.}, C7:233, 1999.

\bibitem{DGP98}
H~G Dosch, T~Gousset, and H~J Pirner.
\newblock {\em Phys. Rev.}, D57:1666, 1998.

\bibitem{L3} 
L3 Collaboration: M.Acciari et al: Phys. Lett. {\bf B408} (1997) 450\\
L3 Note 2280: Submitted to {\em XXIX ICHEP}, Vancouver, 1998\\
 M.Acciari et al. L3 Note 2400 {\em Int. Europhys. Conf. 99}\\
 M.Acciari et al: Phys.Lett. {\bf B453} (1999) 
333 and L3 Note 2404
M.Acciari et al: Phys.Lett. {\bf B436}
(1998)  403 and Phys.Lett. {\bf B447} (1999) 147

\bibitem{OPAL}
OPAL Collaboration: F.W\"ackerle: {\em Proc. XXVIII Int. Symp. on 
Multiparticle Dynamics}, Frascati, 1997\\
 G. Abbiendi et al., to be submitted to Europ. Phys. J. C\\
 K.Ackerstaff et al.: Phys.Lett. {\bf B411}
(1997)  387 and Phys.Lett. {\bf B412} (1997) 225
 OPAL Physics Note PN389 (Preliminary)

\bibitem{ALE}
ALEPH Collaboration: R.Barate et al.: Phys. Lett. {\bf B458} (1999) 152\\
ALEPH 99-038:  EPS-HEP99 Conference, Tampere 

\bibitem{L3C00}
L3 Collaboration L3 Note 2548

\bibitem{OPALC99}
Opal Collaboration: G. Abbiendi et al. To appear in Eur. Phys. J. C


\bibitem{Zeus}
Zeus Collaboration: A. Breitweg et al. hep-ex/9908012

\end{thebibliography}
\section*{Discussion}
{\bf S. Narison},LPMT: { \em I am a bit ennoyed with the $\log( W/m_q^2)$
behaviour of the cross section when $m_q \to 0$. I would have expected a
behaviour like $\log(W/\mu)$ where $\mu$ is a subtraction point. What is the
meaning of $m_q$ for light quarks ?}\\
H.G. Dosch: For real photons the box diagram has an infrared singularity which
is regularized by the quark mass. The same happens for the pomeron
exchange of real photon photon scattering. Here the IR singularity
leads to $1/m_q^4$ ! For light quarks an IR regularization procedure has to be
introduced which at the present state of the art is model dependent.

{\bf A. Brandenburg}: {\em Could you explain in more detail how you obtained
the quark mass that regularized the photon wave function? What is the physical
interpretation? Does it have anything to do with the mass parameter in the
QCD Lagrangian?}\\
H.G. Dosch: The regularization of the photon wave function by an effective
mass was motivated by model investigations~\ct{DGP98}. The values for the
regularizing masses for light quarks are given in equation 9 and are close to
"constituent masses". They were obtained in the following way: The second
derivative of the lowest order two point function for the vector current was
compared with the corresponding phenomonological expression, consisting of the
$\rho-$pole and a continuum. The theoretical expression contains the quark mass
as only free parameter and it was fixed by equating the two expressions. For
$Q^2 >1$ the agreement could not be improved by a finite quark mass and there
the Lagrangian mass can be taken, but it is of little influence anyhow. For
heavy quarks the Lagrangian quark mass at the
scale $Q^2/4$ is a reliable regularizator. For a detailed discussion I refer
to reference \ct{DGP98}.

\end{document}